
\def\today{\ifcase\month\or January\or February\or March\or April\or May\or
June\or July\or August\or September\or October\or November\or December\fi
\space\number\day, \number\year}
%
%
\newcount\notenumber

\def\note{\global\advance\notenumber by 1 \footnote{$^{\the\notenumber}$}}
%
%
\newif\ifsectionnumbering
\newcount\eqnumber
\def\cleareqnumber{\eqnumber=0}
\def\numbereq{\global\advance\eqnumber by 1
\ifsectionnumbering\eqno(\the\secnumber.\the\eqnumber)\else\eqno(\the\eqnumber)
\fi}
\def\eqalinno{{\global\advance\eqnumber by 1}
\ifsectionnumbering(\the\secnumber.\the\eqnumber)\else(\the\eqnumber)\fi}
\def\name#1{\ifsectionnumbering\xdef#1{\the\secnumber.\the\eqnumber}
\else\xdef#1{\the\eqnumber}\fi}
\def\nosectionnumbering{\sectionnumberingfalse}
\sectionnumberingtrue
%
%
\newcount\refnumber

\immediate\openout1=refs.tex
\immediate\write1{\noexpand\frenchspacing}
\immediate\write1{\parskip=0pt}
\def\ref#1#2{\global\advance\refnumber by 1%
[\the\refnumber]\xdef#1{\the\refnumber}%
\immediate\write1{\noexpand\item{[#1]}#2}}
\def\tie{\noexpand~}

%
%
\font\twelvebf=cmbx10 scaled \magstep1
\newcount\secnumber

\def\newsection#1.{\ifsectionnumbering\cleareqnumber\else\fi%
	\global\advance\secnumber by 1%
	\bigbreak\bigskip\par%
	\line{\twelvebf \the\secnumber. #1.\hfil}\nobreak\medskip\par}
%
%
%
\def \sqr#1#2{{\vcenter{\vbox{\hrule height.#2pt
	\hbox{\vrule width.#2pt height#1pt \kern#1pt
		\vrule width.#2pt}
		\hrule height.#2pt}}}}

%
%
%
\newdimen\fullhsize
\def\fiddle{\fullhsize=6.5truein \hsize=3.2truein}
\def\fullline{\hbox to\fullhsize}
\def\mkhdline{\vbox to 0pt{\vskip-22.5pt
	\fullline{\vbox to8.5pt{}\the\headline}\vss}\nointerlineskip}
\def\mkftline{\baselineskip=24pt\fullline{\the\footline}}
\let\lr=L \newbox\leftcolumn
\def\twocolumns{\fiddle
	\output={\if L\lr \global\setbox\leftcolumn=\columnbox
		\global\let\lr=R \else \doubleformat \global\let\lr=L\fi
		\ifnum\outputpenalty>-20000 \else\dosupereject\fi}}
\def\doubleformat{\shipout\vbox{\mkhdline
		\fullline{\box\leftcolumn\hfil\columnbox}
		\mkftline} \advancepageno}
\def\columnbox{\leftline{\pagebody}}
\nosectionnumbering
\magnification=1200
\def\pr#1 {Phys. Rev. {\bf D#1\tie }}
\def\pe#1 {Phys. Rev. {\bf #1\tie}}
\def\pl#1 {Phys. Lett. {\bf #1B\tie }}
\def\prl#1 {Phys. Rev. Lett. {\bf #1\tie }}
\def\np#1 {Nucl. Phys. {\bf B#1\tie }}
\def\ap#1 {Ann. Phys. (NY) {\bf #1\tie }}
\def\cmp#1 {Commun. Math. Phys. {\bf #1\tie }}
\def\imp#1 {Int. Jour. Mod. Phys. {\bf A#1\tie }}
\def\mpl#1 {Mod. Phys. Lett. {\bf A#1\tie}}
\def\tie{\noexpand~}

\parskip=15pt plus 4pt minus 3pt
\headline{\ifnum \pageno>1\it\hfil String Quantization in
Curved Spacetimes
	$\ldots$\else \hfil\fi}
\font\title=cmbx10 scaled\magstep1
\font\tit=cmti10 scaled\magstep1
\footline{\ifnum \pageno>1 \hfil \folio \hfil \else
\hfil\fi}
\raggedbottom
\rightline{\vbox{\hbox{THES-TP-09/94}
\hbox{CTP-TAMU-14/94}\hbox{ACT-42/94}}}
\vfill
\centerline{\title STRING QUANTIZATION IN CURVED
SPACETIMES:}
\centerline{\title NULL STRING APPROACH}
\vfill
\centerline{\title H. J. DE VEGA$^{(a,b)}$,
I. GIANNAKIS$^{(c,d)}$ and A. NICOLAIDIS$^{(e)}$}
\medskip
\centerline{$^{(a)}${\tit LPTHE,
Laboratoire Associ\'e au CNRS UA 280, Universit\'e Paris VI}}
\centerline{\tit F-75252 Paris, France}
\medskip
\centerline{$^{(b)}${\tit Isaac Newton Institute, Cambridge, CB3 0EH, United
Kingdom}}
\medskip
\centerline{$^{(c)}${\tit Center for Theoretical Physics,
Texas A{\&}M
University}}
\centerline{\tit College Station, TX 77843-4242, USA}
\medskip
\centerline{$^{(d)}${\tit Astroparticle Physics Group,
Houston Advanced
Research Center (HARC)}}
\centerline{\tit The Mitchell Campus, Woodlands, TX 77381, USA}
\medskip
\centerline{$^{(e)}${\tit Theoretical Physics Department,
University of Thessaloniki}}
\centerline{\tit Thessaloniki 54006, Greece}
\vfill
\centerline{\title Abstract}
\bigskip
{\narrower\narrower
We study quantum strings in strong gravitational fields. The relevant
small parameter is $g=R_c{\sqrt T_0}$, where $R_c$ is the curvature
of the spacetime and $T_0$ is the string tension. Within our systematic
expansion we obtain to zeroth order the null string (string with
zero tension), while the first order correction incorporates the
string dynamics. We apply our formalism to quantum null strings
in de Sitter spacetime. After a reparametrization of the world-sheet
coordinates, the equations of motion are simplified. The quantum algebra
generated by the constraints is considered, ordering the momentum
operators to the right of the coordinate operators. No critical
dimension appears. It is anticipated however that the conformal
anomaly will appear when the first order corrections proportional
to $T_0$, are introduced. \par}
\vfill\vfill\break

Classical and quantum string propagation in curved spacetimes
is a very important subject. The investigations in this domain
are relevant for the physics of quantum gravitation as well as for the
understanding of the cosmic string models in cosmology.

Strings are characterized by an energy scale
${\sqrt T_0}$ ($T_0$ is the string tension). The frequencies of the
string modes are proportional to $T_0$ and the length of the
string scales with ${1\over {\sqrt T_0}}$. The gravitational field
provides another length scale, the curvature radius of the spacetime
$R_c$. For a string moving in a gravitational field a useful
parameter is the dimensionless constant $g=R_c{\sqrt T_0}$.
Large values of $g$ imply weak gravitational field ( the
metric does not change appreciably over distances of the order
of the string length). We may reach large values of $g$ by
letting $T_0 \to {\infty}$. In this limit the string shrinks to a
point and a suitable expansion (small string oscillations around the
center of mass of the string) has been proposed in ref \ref{\vega}{
H. J. de Vega and N. S\'anchez \pl197 (1987), 320;
For a review see: N. S\'anchez, {\it String Quantum Gravity}, and
H. J. de Vega, {\it Strings in Curved Space-Times},
in N. S\'anchez, ed, {\it ``String Quantum Gravity and Physics at the
Planck scale''} (Erice, Italy, June 1992) World Scientific.}.
In the opposite
limit, small values of $g$, we encounter strong gravitational fields
and it is appropriate to consider $T_0 \to 0$.
In ref
\ref{\nikola}{H. J. de Vega and A. Nicolaidis, \pl295 (1992), 214.}
a systematic expansion in terms of the string tension has been presented.
To zeroth order we obtain the null string
\ref{\shild}{A. Schild, \pr 16 (1977), 1722;
A. Karlhede and U. Lindstr\"om, {\sl Class. Quantum Grav.} {\bf 3}
(1986), L73.}, every point of the string moves independently along a
null geodesic. In this letter we study the quantization of the null
string in curved spacetime, notably in a de Sitter geometry.

Let us summarize the main results of ref. [\nikola]. The limit
$T_0 \to 0$ cannot be reached using the Nambu-Goto action for
the strings. Following the analogous massless particle case we are led
to a reformulated Lagrangian [\nikola]
$$
L={1\over 4{\lambda}}[{\dot X^{\mu}}{\dot X^{\nu}}G_{\mu\nu}(X)
-c^2{X'^{\mu}}{X'^{\nu}}G_{\mu\nu}(X)] \numbereq\name\eqghio
$$
where $c=2{\lambda}T_0$ is the world-sheet speed of light
( a dot and a prime denote respectively differentiation
with respect to the world-sheet time and space variables,
$\tau$ and $\sigma$). The string equations of motion read
$$
{\ddot X^{\mu}}-c^2 X''^{\mu}
+{\Gamma^{\mu}_{\kappa\lambda}}({\dot X^{\kappa}}{\dot
X^{\lambda}}-c^2 X'^{\kappa} X'^{\lambda})=0
\numbereq\name{\eqhatzid}
$$
supplemented by the constraints
$$
{\dot X^{\mu}}X'^{\nu}G_{\mu\nu}=0 \qquad
{\dot X^{\mu}}{\dot X^{\nu}}G_{\mu\nu}
+c^2{X'^{\mu}}{X'^{\nu}}G_{\mu\nu}=0
\numbereq\name{\eqbouris}
$$
A systematic expansion in powers of $c$ is feasible now [\nikola].
If we write
$$
X^{\mu}({\sigma}, {\tau})=A^{\mu}({\sigma}, {\tau})
+c^2B^{\mu}({\sigma}, {\tau})+ \cdots \numbereq\name{\eqhuop}
$$
where the dots indicate
higher powers of $c^2$ and substitute this form of $X^\mu$ in
(\eqhatzid) and (\eqbouris) we find that to zeroth order in $c$
the string dynamics is given by the equations of motion
$$
{\ddot A^{\mu}}+{\Gamma^{\mu}_{\kappa\nu}}
{\dot A^{\kappa}}{\dot A^{\nu}}=0 \numbereq\name{\eqargu}
$$
and the constraints
$$
\eqalignno{
{\dot A^{\mu}}{\dot A^{\nu}}G_{\mu\nu}&=0&{\global \advance
\eqnumber by 1}(\the\eqnumber a)\name{\eqatma}\cr
{\dot A^{\mu}}A'^{\nu}G_{\mu\nu}&=0&(\the\eqnumber b)\cr}
$$
$A^{\mu}({\sigma}, {\tau})$ represents a collection of
massless particles moving independently along null geodesics. The
only reminiscence from the string is the constraint,
eq. (\eqatma b), which requires the velocity to be perpendicular
to the string.
The next order correction $B^{\mu}(\sigma,\tau)$ obeys the
following equation of motion
$$
{\ddot B^{\mu}}+{\Gamma^{\mu}_{\kappa\lambda}}(
{\dot A^{\lambda}}{\dot B^{\kappa}}+{\dot A^{\kappa}}
{\dot B^{\lambda}})+{\Gamma^{\mu}_{\kappa\lambda,\nu}}
{\dot A^{\kappa}}{\dot A^{\lambda}}B^{\nu}=
A''^{\mu}+{\Gamma^{\mu}_{\kappa\lambda}}A'^{\kappa}
A'^{\lambda} \numbereq\name{\eqrezil}
$$
supplemented with the constraints
$$
\eqalign{
&2{\dot A^\mu}{\dot B^\nu}G_{\mu\nu}+{\dot A^\mu}{\dot A^\nu}
B^{\rho}G_{\mu\nu,\rho}+A'^{\mu}A'^{\nu}G_{\mu\nu}=0\cr
&{\dot B^\mu}A'^{\nu}G_{\mu\nu}+{\dot A^{\mu}}B'^{\nu}G_{\mu\nu}
+{\dot A^\mu}A'^{\nu}B^{\rho}G_{\mu\nu,\rho}=0 \cr}
\numbereq\name{\eqgout}
$$
where ${\Gamma^{\mu}_{\kappa\lambda,\rho}}$ and $G_{\mu\nu,\rho}$
indicate the derivatives with respect to $A^{\rho}$.

We would like to study the quantization of the null string in
de Sitter spacetime. The line element is defined as
$$
ds^2=C^2(X_0)(dX_{0}^2-dX_{1}^2-dX_{2}^2-dX_{3}^2)
\numbereq\name{\eqwobut}
$$
with
$$
C(X_0)={R_{0}\over X_0} \numbereq\name{\eqridik}
$$
where $X_0$ is the conformal
time and $R_0$ is the scale factor. The Lagrangian for a null
string propagating in (\eqwobut) takes the form
$$
L=C^2(A_0)({\dot A}_{0}^2-{\dot A}_{1}^2-{\dot A}_{2}^2
-{\dot A}_{3}^2)
\numbereq\name{\eqgrounion}
$$
The equations of motion, using also the constraints, provide
$$
\eqalignno{
C^2(A_0){\dot A}_i&=P_i({\sigma}) \qquad i=1,2,3&{\global\advance
\eqnumber by 1}(\the\eqnumber a)\name{\eqgoufi} \cr
C^2(A_0){\dot A}_0&=P_0({\sigma})& (\the\eqnumber b)\cr}
$$
Using eq. (\eqridik), we obtain from eq. (\eqgoufi b)
$$
A_0({\sigma}, {\tau})={{{\overline A}_0R_{0}^2}\over {R_{0}^2-
{\overline A}_{0}P_0{\tau}}} \numbereq\name{\eqwlko}
$$
with ${\overline A}_0=A_0({\sigma}, {\tau}=0)$.
Eq. (\eqgoufi a) gives then
$$
A_i({\sigma}, {\tau})-{\overline A}_i
={{P_i{\overline A}_0^2{\tau}}\over {R_{0}^2-
{\overline A}_{0}P_0{\tau}}} \numbereq\name{\eqmuchac}
$$
 with ${\overline A}_i=A_i({\sigma}, {\tau}=0)$. The constraints
(\eqatma) tie the initial shape and momentum of the
string
$$
\eqalignno{
{\overline A'}_0P_0&={\sum_{i=1}^3}{\overline A'}_iP_i&
{\global\advance\eqnumber by 1} (\the\eqnumber a)\name{\equrinp} \cr
P_{0}^2({\sigma})&={\sum_{i=1}^3}P_{i}^2(\sigma)&(\the\eqnumber b)\cr}
$$
Notice that the relationship
$$
A_i-{\overline A}_i={P_{i}\over
P_{0}}(A_0-{\overline A}_0) \numbereq\name{\eqrcnsd}
$$
holds, i.e in terms of the cosmic
variables we obtain straight lines for the individual particles
of the null string. The constraints retain their form under the
reparametrization ${\tau}=f({\tilde {\tau}}, {\tilde {\sigma}})$
and ${\sigma}=g({\tilde {\sigma}})$ with $f$ and $g$ arbitrary
functions. We may use this reparametrization freedom in order to
simplify the equations of motion. Choosing
$$
{\tau}={R_{0}^2{\tilde {\tau}}\over {{\overline A}_0P_0
{\tilde {\tau}}+{\overline A}_{0}^2}} \numbereq\name{\eqrfmds}
$$
we obtain (after dropping tildes)
$$
\eqalignno{
A_{\mu}&=P_{\mu}{\tau}+
{\overline A}_{\mu} \qquad {\mu}=0,1,2,3
&{\global\advance\eqnumber by 1}(\the\eqnumber a)\name{\eqgoliah} \cr
P_{\mu}^2&=0& (\the \eqnumber b)\cr
{\overline A'}_{\mu}&P^{\mu}=0& (\the \eqnumber c)\cr}
$$
Due to the periodicity in the
${\sigma}$-direction, since we deal with closed strings,
we can expand $P^{\mu}({\sigma})$ and ${\overline
A^{\mu}}({\sigma})$ in Fourier series as follows
$$
P^{\mu}({\sigma})={\sum_n}\;p^{\mu}_n\;{\exp(in{\sigma})} \qquad
{\overline A^{\mu}}({
\sigma})={\sum_n}\;x^{\mu}_n\;{\exp (in{\sigma})} \numbereq\name{\eqstella}
$$
Similarly the constraints can be expanded in Fourier series
$$
P^{\mu}P_{\mu}={\sum_{\kappa}}H_{\kappa}\;{\exp (i{\kappa}{\sigma})}
\qquad P_{\mu}{\overline A'^{\mu}}
={\sum_{\kappa}}G_{\kappa}\;{\exp (i{\kappa}{\sigma})}
\numbereq\name{\eqaggelos}
$$
By substituting the expressions for the
$P^{\mu}({\sigma})$ and ${\overline A}({\sigma})$
in the above relations we find
$$
H_{\kappa}={\sum_n}\;p_{k-n}\;p_{n} \qquad
G_{\kappa}=i{\sum_n}n\;p_{k-n}\;x_{n} \numbereq\name{\eqzarko}
$$
The string coordinate $\overline A^{\mu}({\sigma})$
and the conjugate momentum
$P^{\mu}({\sigma})$ satisfy the following Poisson brackets
$$
{\lbrace P^{\mu}({\sigma}), {\overline A^{\nu}}({\sigma'}) \rbrace}=
{\delta}({\sigma}-{\sigma'})G^{\mu \nu} \numbereq\name{\eqrfho}
$$
while for their Fourier modes $p_{n}^{\mu}$ and $x_{n}^{\mu}$
we obtain
$$
{\lbrace p_{n}^{\mu}, x_{m}^{\nu} \rbrace}
=G^{\mu \nu}{\delta}_{m+n}
$$
Using the above relations we can calculate the Poisson brackets of the moments
$H_{\kappa}$ and $G_{\kappa}$ of the constraints.
With a redefinition $G_{\kappa} \to -G_{\kappa}$
we find the following algebra (the analogue of the
{\it Vir}$\times${\it Vir} in the tensionful string)
$$
{\lbrace G_n, G_m \rbrace}=i(n-m)G_{n+m}, \quad {\lbrace G_n, H_m \rbrace}=
i(n-m)H_{n+m}, \quad {\lbrace H_n, H_m \rbrace}=0 \numbereq\name{\eqroi}
$$
The constraints $G_n$ and $H_n$ generate reparametrizations in
${\sigma}$ and ${\tau}$, respectively.
The algebra of the constraints of the null string arises as the Inonu-Wigner
contraction \ref{\freo}{E. Inonu and E. Wigner, {\it Proc. Nat. Acad. Sci.
(US)} {\bf 39} (1953), 510.}
$(T_0 \to 0)$ of the algebra of the constraints of the
tensile string. This can be seen by writting the constraint algebra
of the tensile string in terms of the modes ${\tilde H}_m$, ${\tilde
G}_n$ of the constraints $PX'=P^2+{T_0}^2X'^2=0$.
The modes of the constraints of the tensile string
${\tilde G_n}, {\tilde H_n}$ are given in
terms of the modes of the constraints of the null string $G_n, H_n$
$$
{\tilde G_n}=G_n, \qquad {\tilde H_n}=H_n+{T_0}^2{\sum_{\kappa}}({\kappa}
-n){\kappa}x_{(n-{\kappa})}x_{\kappa} \numbereq\name{\eqrtip}
$$
The algebra then takes the form
$$
\eqalign{
{\lbrace {\tilde G}_n, {\tilde G}_m \rbrace}&=i(n-m)\;{\tilde G}_{n+m},
\quad {\lbrace {\tilde G}_n, {\tilde H}_m \rbrace}=
i(n-m){\tilde H}_{n+m}\cr
{\lbrace {\tilde H}_n, {\tilde H}_m \rbrace}
&=4i{T_0}^2(n-m)\;{\tilde G}_{m+n}\cr}
\numbereq\name{\eqroi}
$$
which clearly in the limit $T_0 \to 0$ reduces to the null algebra.

We may now quantize the null string by replacing the Poisson bracket
by the commutator
$$
{\lbrack p^{\mu}_n, x^{\nu}_m \rbrack}=-iG^{\mu\nu}{\delta_{m+n}}
\numbereq\name{\eqfcoi}
$$
The passage from the classical
domain to the quantum one offers the possibility for the emergence of central
terms in the algebra. The algebra will take the form
$$
\eqalign
{&{\lbrack G_n, G_m \rbrack}=(n-m)G_{n+m}+A(n){\delta}_{n+m}, \quad
{\lbrack G_n, H_m \rbrack}=
(n-m)H_{n+m}+B(n){\delta}_{n+m}, \quad \cr
&{\lbrack H_n, H_m \rbrack}=C(n){\delta}_{n+m} \cr} \numbereq\name{\eqpani}
$$
In order to specify the form of the anomaly we make use of the Jacobi identity
for the algebra. We find that the most general form of the anomaly is given
by
$$
A(n)={\alpha}n^3+{\beta}n, \qquad B(n)={\gamma}n^3+{\delta}n,
\qquad C(n)=0 \numbereq\name{\eqdragan}
$$
where $\alpha$, $\beta$, $\gamma$ and $\delta$
are constants which need to be
determined.
The actual calculation depends crucially upon the ordering of
the operators \ref{\liz}{F. Lizzi, B. Rai, G. Sparano
and A. Srivastava, \pl182 (1986), 326; F. Lizzi, \mpl9 (1994), 1465.},
\ref{\gam}{J. Gamboa, C. Ram\'{\i}rez and M. Ruiz-Altaba, \np 338
(1990), 143.}, \ref{\theord}{J. Isberg, U. Lindstr\"om, B. Sundborg and
G. Theodoridis, \np 411 (1994), 122.}.
Given our Lagrangian eq. (\eqghio) with $c=0$, the vacuum
$|0>$ satisfies the condition
$$
p^{\mu}_n|0>=0 \numbereq\name{\eqrvbs}
$$
We order then all the $p_n$ to the right of the $x_n$. We arrive at
similar conclusions by looking at the complete tensile
theory and considering the limit $c \to 0$.
The tensile string is quantized using harmonic oscillator operators,
representing right and left movers. The operators which annihilate
the vaccum state in the limit $c \to 0$ are reduced to momentum
operators [\theord]. With the adopted ordering,
it is relatively easy to evaluate the quantum algebra generated by the
constraints. The quantum algebra {\bf does not contain central terms},
namely we find that $\alpha=\beta=\gamma=\delta=0$ and
therefore a quantum null string may exist in any dimension without anomalies.
We anticipate however that a critical dimension will appear,
as soon as the first order correction $B^{\mu}({\sigma}, {\tau})$,
proportional to $T_0$, is introduced (work in progress).

Summarizing, we presented a consistent framework within which
it is feasible to study string quantization in strong gravitational
fields. We have shown that for the emergence of the conformal anomaly,
responsible is the dimensonful string tension. Historical precedent
(Higgs mechanism) may allow us in contemplation for a mechanism
generating string tension and at the same time respecting the
symmetries.
\bigbreak
\line{\bf Acknowledgements.\hfil}
We would like to acknowledge useful discussions with our
colleagues, HdV with N. S\'anchez and G. Veneziano, IG with M. Evans and
A. Polychronakos and AN with J. Iliopoulos and K. Tamvakis.
\nobreak\bigskip

\immediate\closeout1
\bigbreak\bigskip

\line{\bf References.\hfil}
\nobreak\medskip\vskip\parskip

\input refs

\vfill\end

\bye